\documentclass[aps,prd,nofootinbib,amsmath,amssymb,superscriptaddress,twocolumn,10pt]{revtex4}
\usepackage{graphicx}
\usepackage{dcolumn}
\usepackage{bm}
\usepackage{amssymb}
\usepackage{latexsym}
\usepackage{booktabs}
\usepackage{amsmath}
\usepackage{multirow}
\usepackage[colorlinks=true, linkcolor=red, citecolor=blue]{hyperref}
\usepackage{color}
\newcommand{\be}{\begin{equation}}
\newcommand{\ee}{\end{equation}}
\newcommand{\bq}{\begin{eqnarray}}
\newcommand{\eq}{\end{eqnarray}}

\begin{document}

\title{Redshift drift exploration for interacting dark energy}

\author{Jia-Jia Geng}
\affiliation{Department of Physics, College of Sciences, Northeastern University, Shenyang
110004, China}
\author{Yun-He Li}
\affiliation{Department of Physics, College of Sciences, Northeastern University, Shenyang
110004, China}
\author{Jing-Fei Zhang}
\affiliation{Department of Physics, College of Sciences, Northeastern University, Shenyang
110004, China}
\author{Xin Zhang}
\email{zhangxin@mail.neu.edu.cn} \affiliation{Department of Physics, College of Sciences,
Northeastern University, Shenyang 110004, China}
\affiliation{Center for High Energy Physics, Peking University, Beijing 100080, China}

\begin{abstract}
By detecting redshift drift in the spectra of Lyman-$\alpha$ forest of distant quasars, Sandage-Loeb (SL) test directly measures the expansion of the universe, covering the ``redshift desert" of $2 \lesssim z \lesssim5$. Thus this method is definitely an important supplement to the other geometric measurements and will play a crucial role in cosmological constraints. In this paper, we quantify the ability of SL test signal by a CODEX-like spectrograph for constraining interacting dark energy. Four typical interacting dark energy models are considered:
(\romannumeral1) $Q=\gamma H\rho_c$, (\romannumeral2) $Q=\gamma H\rho_{de}$, (\romannumeral3) $Q=\gamma H_0\rho_c$, and (\romannumeral4) $Q=\gamma H_0\rho_{de}$.
The results show that for all the considered interacting dark energy models, relative to the current joint SN+BAO+CMB+$H_0$ observations, the constraints on $\Omega_m$ and $H_0$ would be improved by about 60\% and 30--40\%, while the constraints on $w$ and $\gamma$ would be slightly improved, with a 30-yr observation of SL test.
We also explore the impact of SL test on future joint geometric observations. In this analysis, we take the model with $Q=\gamma H\rho_c$ as an example, and simulate future SN and BAO data based on the space-based project WFIRST. We find that in the future geometric constraints, the redshift drift observations would help break the geometric degeneracies in a meaningful way, thus the measurement precisions of $\Omega_m$, $H_0$, $w$, and $\gamma$ could be substantially improved using future probes. 

\end{abstract}

 \maketitle

\section{Introduction}

Since interactions are ubiquitous in nature, it is rather natural to imagine that dark energy might directly interact with cold dark matter. Actually, that there is no any direct interaction between dark energy and dark matter is an additional, strong assumption. Currently, one of the most important missions in the field of dark energy research is to provide positive/negative evidence for (i.e., certify/falsify) the scenario of interacting dark energy in the light of observational data. Synthetically using the measurements of expansion history and growth of structure to consistently test the scenario is a fairly important way. However, for the scenario of interacting dark energy, it is difficult to finely test models using the measurements of growth of structure due to some complexness, such as the
diversity of the construction of covariant 4-vector interaction, the large-scale gravity instability, and the lack of  abundant, highly accurate data of growth of structure. Although the progress has been made in this aspect since the parametrized post-Friedmann theoretical framework for interacting dark energy was proposed and applied \cite{PPF1,PPF2}, obstacle due to other factors still exists. Under such circumstance, in this work, we only consider to use the geometric measurements to constrain the interacting dark energy models; in particular, we focus on the future redshift drift data.

As a purely geometric measurement, Sandage-Loeb (SL) test will be crucial to probe the ``redshift desert" ($2 \lesssim z \lesssim5$) by directly measuring the expansion of the universe. It was firstly proposed by Sandage~\cite{sandage} to directly measure the variation of redshift of distant sources. Then Loeb~\cite{loeb} found a realistic way of detecting redshift drift in the spectra of Lyman-$\alpha$ forest of distant quasars (QSOs). The 39-meter European Extremely Large Telescope (E-ELT) in built was equipped with a high-resolution spectrograph called CODEX (COsmic Dynamics Experiment), which was designed to achieve this goal. Cosmological constraints with SL test have been studied by numerous works~\cite{sl1,sl2,sl3,sl4,sl5,sl6,sl7,Darling,Zhang21}, some of which simulated 240 or 150 quasars to be observed. However, as pointed in Ref.~\cite{Liske}, only about 30 quasars are bright enough or lying at high enough redshift for the actual observation, based on a Monte Carlo simulation using a telescope with a spectrograph like CODEX.  Besides, as far as we know, in the most existing papers, the best-fit $\Lambda$CDM model to current data is chosen as the fiducial model, based on which SL test data are simulated. Thus, when these SL test data are further combined with other actual data to constrain dark energy models, tension between the simulated SL data and other actual data may occur. This is inappropriate and may not give convincing conclusion on the impact of future SL test data on cosmological constraints.

To avoid inconsistency in data, in our previous work~\cite{msl1}, we suggested that the best-fit model in study to current actual data is chosen as the fiducial model in simulating 30 mock SL test data. To give a typical example, we only focused on the dark energy model with constant $w$ (referred to as $w$CDM model). In our recent work~\cite{msl2}, we extended the discussion to time-evolving dark energy model and explored the impact of SL test data on dark energy constraints in the future geometric measurements.

Though in Ref.~\cite{msl1}, a preliminary SL test analysis has been made for two simple interacting dark energy models, a synthetical analysis in depth in quantifying the impact of future redshift drift data on testing different types of interacting dark energy models is still absent. This paper will provide such an analysis.
In this paper, we will quantify the constraining power of future SL test data on different interacting dark energy models, and show how the SL test impacts on the parameter estimation.

For interacting dark energy models, the energy balance equations for dark energy and cold dark matter are
\begin{equation}
\label{eqq1}
 \dot{\rho}_{de}+3H\rho_{de}(1+w)=-Q,
\end{equation}
\begin{equation}
\label{eqq2}
 \dot{\rho}_c+3H \rho_c=Q,
\end{equation}
where $\rho_{de}$ and $\rho_c$ are the background energy densities of dark energy and cold dark matter, respectively. The Hubble parameter
$H=\dot{a}/a$ describes the expansion rate of the universe and the interacting term $Q$ describes the energy transfer rate between dark energy and dark matter densities.
We consider four typical interacting dark energy models: (\romannumeral1) $Q=\gamma H\rho_c$, with which the model is called the I$w$CDM1 model, (\romannumeral2) $Q=\gamma H\rho_{de}$, with which the model is called the I$w$CDM2 model, (\romannumeral3) $Q=\gamma H_0\rho_c$, with which the model is called the I$w$CDM3 model, and
(\romannumeral4) $Q=\gamma H_0\rho_{de}$, with which the model is called the I$w$CDM4 model.
Here $\gamma$ is the dimensionless coupling parameter, and the equation-of-state parameter $w$ is considered to be a constant in this paper.

In fact, when the CODEX experiment is ready to deliver its redshift drift data, other future geometric measurements data will also be available. Therefore, it seems that a more meaningful issue is to ask what role the SL test will play in parameter estimation for interacting dark energy models using the future geometric measurements. To address this issue, we also simulate the future SN and BAO data based on the long-term space-based project WFIRST (Wide-Field Infrared Survey Telescope).
 It is quite common that program in built is changed or postponed. Besides, because science and technology develop rapidly, it is hard to well and truly quantify the percentages of the parameter estimation by SL test data. We only take WFIRST as an example. Owing to the fact that it covers the redshift desert ($2 \lesssim z \lesssim5$), SL test is a valuable supplementary to other geometric measurements. Moreover, because of the different degeneracy orientations of SL test and other geometric measurements, it is very credible that SL test can effectively break the strong degeneracy in other geometric measurements and greatly improve the corresponding cosmological constraint results. It's necessary to point out that this analysis does not mean the necessity of combining the data from WFIRST and CODEX in the future, and that we merely explore the ability of redshift drift to break degeneracy in other geometric measurements and to improve the accuracy of cosmological constraints.

\section{Methodology}

Our procedure is as follows. Interacting dark energy models are first constrained by using the current joint SN+BAO+CMB+$H_0$ data, and then for each case
the best-fit model is chosen to be the fiducial model in producing the simulated mock SL test data.
The obtained SL test data are thus well consistent with the current data.
Therefore, it is rather appropriate to combine the mock SL test data with the current data
for further constraining the interacting dark energy models. We perform an MCMC likelihood analysis~\cite{cosmomc} to obtain ${\cal O}(10^6)$ samples for each model.


For current data, the most typical geometric measurements are chosen, i.e., the observations of SN, BAO, CMB, and $H_0$.
For the SN data, the SNLS compilation~\cite{snls3} with a sample of 472 SNe is used.
For the BAO data, we consider the $r_s/D_V(z)$ measurements from 6dFGS ($z=0.1$)~\cite{6dF}, SDSS-DR7 ($z=0.35$)~\cite{DR7}, SDSS-DR9 ($z=0.57$)~\cite{DR9}, and
WiggleZ ($z=0.44$, 0.60, and 0.73)~\cite{WiggleZ} surveys.
For the CMB data, we use the Planck distance posterior given by Ref.~\cite{WW}.
As dark energy only affects the CMB through the comoving angular diameter distance to the decoupling epoch (and the late-time ISW effect), the distance information given by the CMB distance posterior is sufficient for the joint geometric constraint on dark energy.
We also use the direct measurement result of the Hubble constant in the light of the cosmic distance ladder from the HST, $H_0=73.8\pm 2.4$ km s$^{-1}$ Mpc$^{-1}$~\citep{Riess2011}.

Next, we briefly review how to simulate the SL test data.
This method is just to directly measure the redshift variation of quasar Lyman-$\alpha$ absorption lines.
The redshift variation is defined as a spectroscopic velocity shift \cite{loeb},
\begin{equation}\label{eq6}
\ \Delta v \equiv \frac{\Delta z}{1+z}=H_0\Delta t_o\bigg[1-\frac{E(z)}{1+z}\bigg],
\end{equation}
where $\Delta t_o$ is the time interval of observation, and $E(z)=H(z)/H_0$ is given by specific dark energy models.

According to the Monte Carlo simulations, the uncertainty of $\Delta v$ expected by CODEX can be expressed as~\cite{Liske}
\begin{equation}\label{eq7}
\sigma_{\Delta v}=1.35
\bigg(\frac{S/N}{2370}\bigg)^{-1}\bigg(\frac{N_{\mathrm{QSO}}}{30}\bigg)^{-1/2}\bigg(
\frac{1+z_{\mathrm{QSO}}}{5}\bigg)^{x}~\mathrm{cm}~\mathrm{s}^{-1},
\end{equation}
where $S/N$ is the signal-to-noise ratio defined per 0.0125 ${\AA}$ pixel, $N_{\mathrm{QSO}}$ is the
number of observed quasars, $z_{\mathrm{QSO}}$ represents their
redshift, and the last exponent $x=-1.7$ for $2<z<4$ and $x=-0.9$ for $z>4$.

To simulate the SL test data, we first constrain the interacting dark energy models by using the current data combination.
The obtained best-fit parameters are substituted into Eq.~(\ref{eq6}) to get the central values of the SL test data.
We choose $N_{\mathrm{QSO}}=30$ mock SL data uniformly distributed among
six redshift bins of $z_{\rm QSO}\in [2, 5]$ and typically take $\Delta t_o=30$ yr in our analysis.
The error bars are computed from Eq.~(\ref{eq7}) with $S/N=3000$.

In order to quantify the power of SL test in future high-precision joint geometric constraints on dark energy, we simulate future SN and BAO data based on the long-term space-based project WFIRST using the method presented in Ref.~\cite{DETF}, and take the interacting dark energy model with $Q=\gamma H\rho_c$ as an example.
We simulate 2000 future SNe distributed in 16 bins over the range $z=0.1$ to $z=1.7$. The observables are apparent magnitudes $m_i=M+\mu(z_i)$, where $M$ represents the absolute magnitude, and $\mu(z_i)$ is the distance modulus. We also include an additional ``near sample" of 500 SNe at $z\approx0.025$.
For future BAO data, we simulate 10000 mock BAO data uniformly distributed among 10 redshift bins of $z\in [0.5,~2]$, with each $\Delta z_i$ centered on the grid $z_i$. The observables are expansion rate $H(z)$ and comoving angular diameter distance $d_A^{co}(z)=d_L(z)/(1+z)$.
For details, we also refer the reader to Ref. \cite{msl2}.

\section{Results and discussion}
\begin{table*}\small
\setlength\tabcolsep{5pt}
\caption{Fit results for the $w$CDM, I$w$CDM1, I$w$CDM2, I$w$CDM3 and I$w$CDM4 models using the
current CMB+BAO+SN+$H_0$ data.}
\label{table1}
\renewcommand{\arraystretch}{1.5}\centering
\begin{tabular}{cccccccccc}
\\
\hline\hline
Parameter  & $w$CDM & I$w$CDM1 & I$w$CDM2 & I$w$CDM3 & I$w$CDM4\\ \hline

$\Omega_bh^2$      & $0.02218^{+0.00025}_{-0.00029}$
                   & $0.02229^{+0.00033}_{-0.00024}$
                   & $0.02232\pm0.00027$
                   & $0.02231\pm0.00028$
                   & $0.02230^{+0.00030}_{-0.00026}$
                   \\

$\Omega_ch^2$      & $0.1201^{+0.0020}_{-0.0022}$
                   & $0.1288^{+0.0070}_{-0.0049}$
                   & $0.1209^{+0.0019}_{-0.0022}$
                   & $0.1215^{+0.0020}_{-0.0023}$
                   & $0.1206^{+0.0021}_{-0.0019}$
                   \\

$w$                & $-1.103\pm0.058$
                   & $-1.136^{+0.062}_{-0.055}$
                   & $-1.152^{+0.064}_{-0.072}$
                   & $-1.156^{+0.070}_{-0.060}$
                   & $-1.149^{+0.063}_{-0.077}$
                   \\

$\gamma$           & ...
                   & $-0.0112^{+0.0054}_{-0.0078}$
                   & $-0.0284^{+0.0218}_{-0.0206}$
                   & $-0.0280^{+0.0194}_{-0.0201}$
                   & $-0.0299^{+0.0215}_{-0.0264}$
                   \\

$\Omega_{m}$       & $0.2844^{+0.0104}_{-0.0093}$
                   & $0.2849^{+0.0114}_{-0.0085}$
                   & $0.2834^{+0.0086}_{-0.0114}$
                   & $0.2825^{+0.0096}_{-0.0100}$
                   & $0.2822^{+0.0102}_{-0.0095}$
                   \\

$H_0$              & $70.74^{+1.26}_{-1.30}$
                   & $72.81^{+1.87}_{-1.67}$
                   & $71.09^{+1.39}_{-1.10}$
                   & $71.34\pm1.33$
                   & $71.15^{+1.27}_{-1.25}$
                   \\
\hline
\end{tabular}
\end{table*}

First, we constrain the $w$CDM model and four typical interacting dark energy models from the current SN+BAO+CMB+$H_0$ data combination, and present the detailed fit results in Table~\ref{table1}.
From this table, one can clearly see that for $w$CDM model, $w<-1$ is preferred at about the 1.8$\sigma$ level, while $w<-1$ is preferred at more than 2.2$\sigma$ level for all the four interacting dark energy models.
For the $w$CDM, I$w$CDM2, I$w$CDM3 and I$w$CDM4 models, $\Omega_c h^2$ can be tightly constrained, and a smaller value is
preferred. But for the I$w$CDM1 model, $\Omega_c h^2$ cannot be well constrained, and a bigger value is more
favored in this case.
The coupling $\gamma$ is tightly constrained in the I$w$CDM1 model, but its constraint is
much weaker in the I$w$CDM2, I$w$CDM3 and I$w$CDM4 models.
For I$w$CDM1 model, $\gamma<0$ is preferred at about 2.1$\sigma$ level, while $\gamma<0$ is slightly favored at about 1.4$\sigma$ level for I$w$CDM2, I$w$CDM3 and I$w$CDM4 models.

\begin{figure}
\begin{center}
\includegraphics[width=8cm]{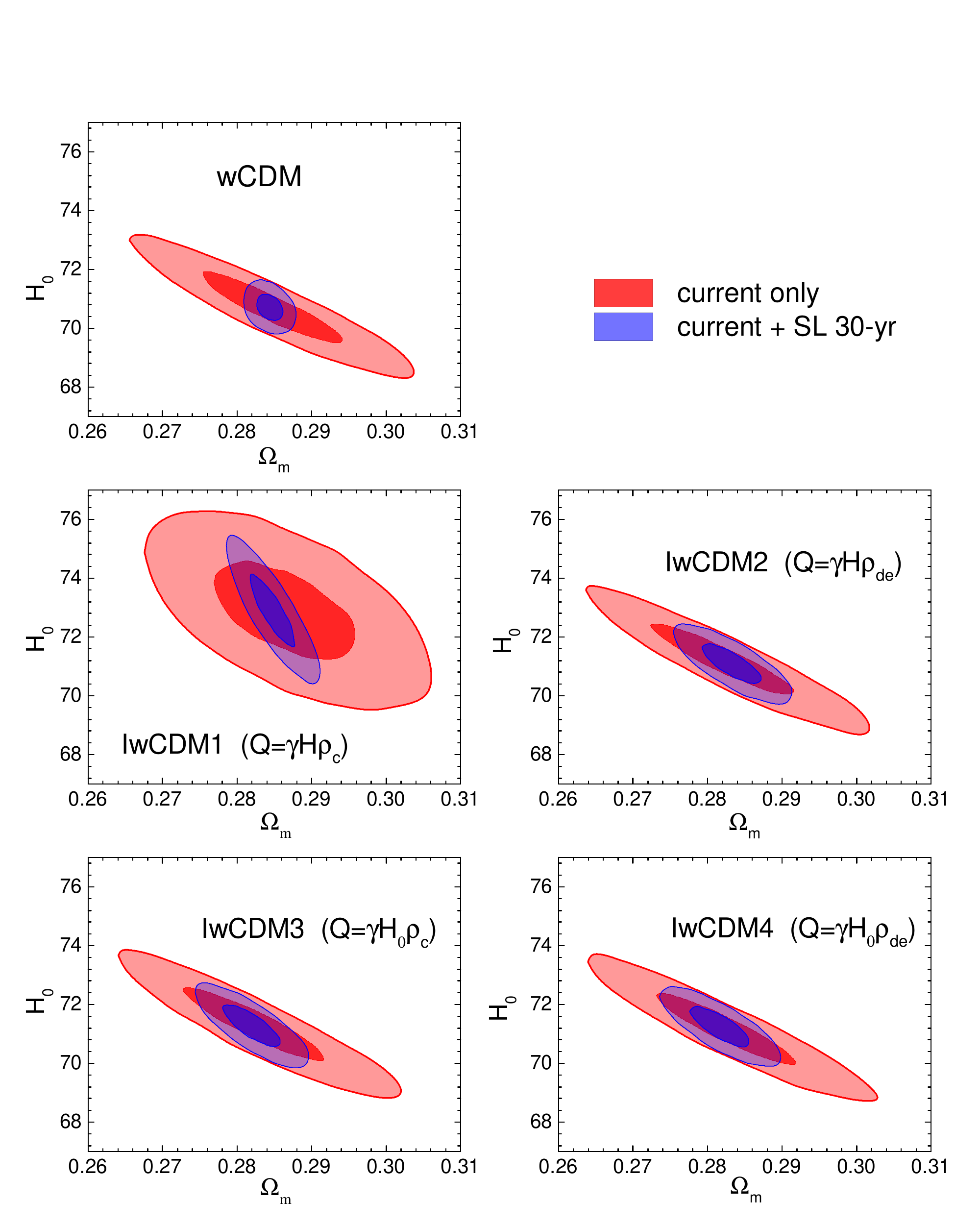}
\end{center}
\caption{Constraints (68.3\% and 95.4\% CL) in the $\Omega_m$--$H_0$ plane for $w$CDM, I$w$CDM1, I$w$CDM2, I$w$CDM3, and I$w$CDM4 models with current only and current+SL 30-yr data.}
\label{fig1}
\end{figure}

\begin{figure}
\begin{center}
\includegraphics[width=8cm]{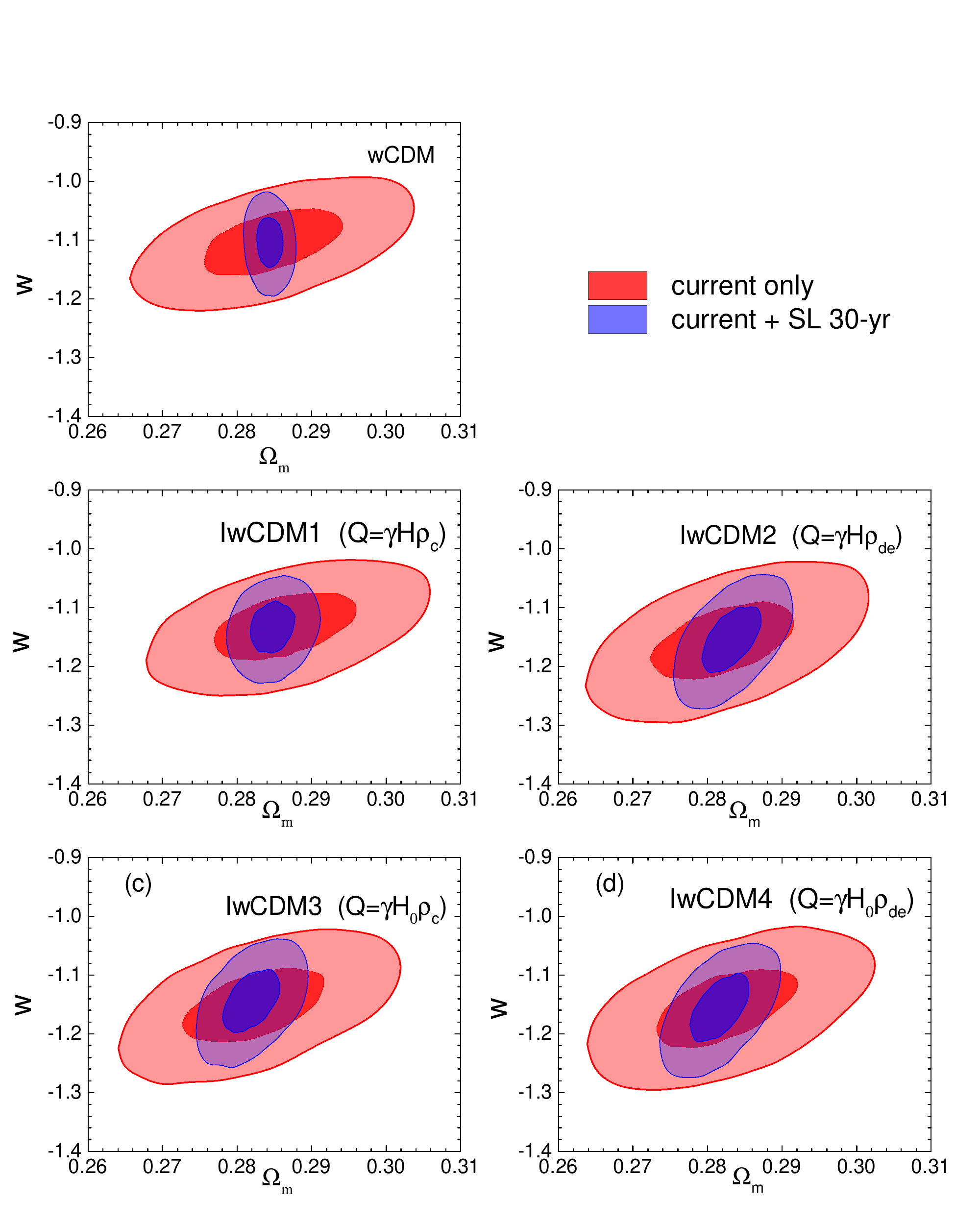}
\end{center}
\caption{Constraints (68.3\% and 95.4\% CL) in the $\Omega_m$--$w$ plane for $w$CDM, I$w$CDM1, I$w$CDM2, I$w$CDM3, and I$w$CDM4 models with current only and current+SL 30-yr data.}
\label{fig2}
\end{figure}

\begin{figure}
\begin{center}
\includegraphics[width=8cm]{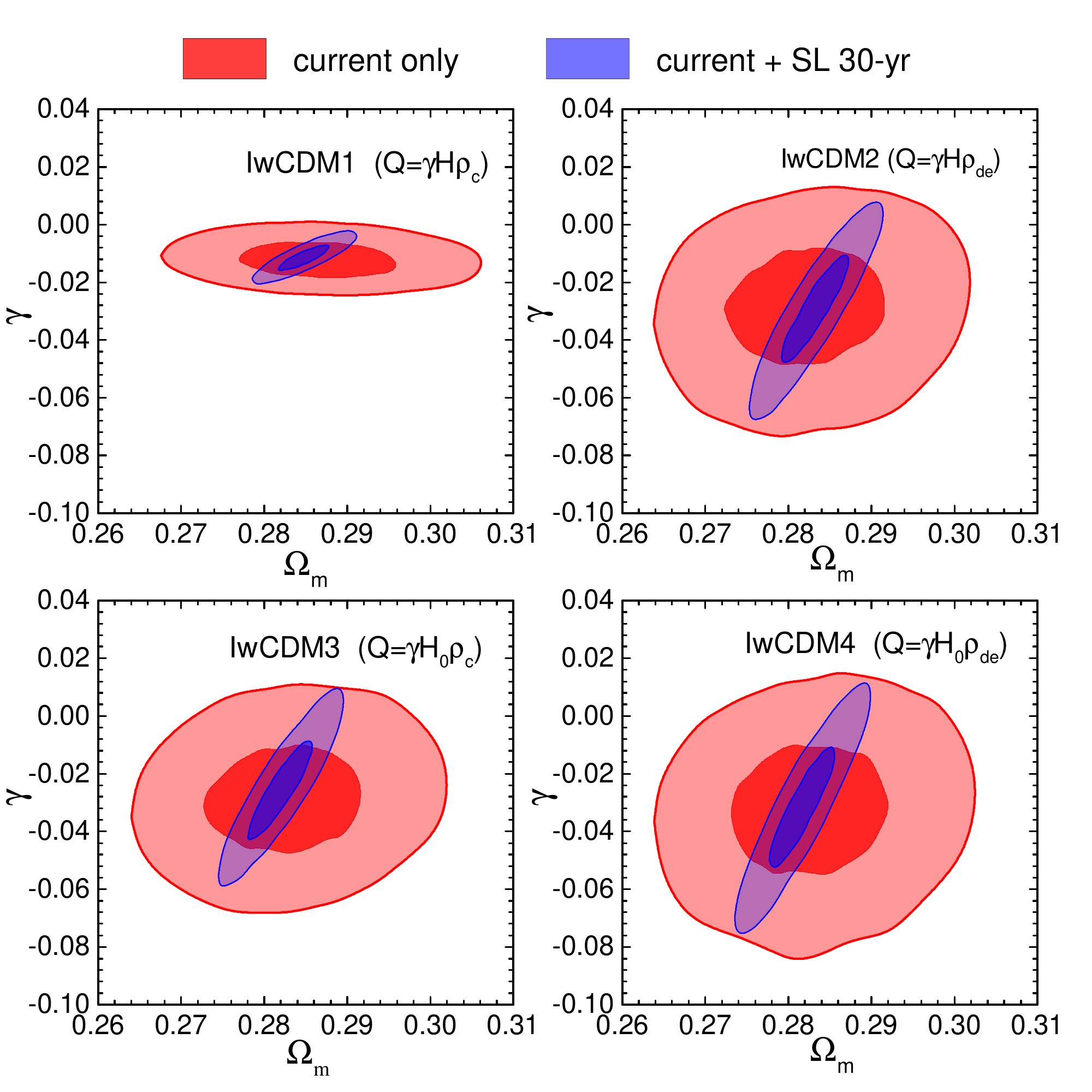}
\end{center}
\caption{Constraints (68.3\% and 95.4\% CL) in the $\Omega_m$--$\gamma$ plane for I$w$CDM1, I$w$CDM2, I$w$CDM3, and I$w$CDM4 models with current only and current+SL 30-yr data.}
\label{fig3}
\end{figure}

Then we combine the simulated 30-yr SL test data with current data and show the constraint results in Figs. \ref{fig1}--\ref{fig3}. To give an intuitive comparison, the results from current only data are also presented.
Figures \ref{fig1} and \ref{fig2} show the joint constraints on the $w$CDM, I$w$CDM1, I$w$CDM2, I$w$CDM3 and I$w$CDM4 models in the $\Omega_m$--$H_0$ and $\Omega_m$--$w$ planes, respectively.
Figure~\ref{fig3} shows the joint constraints on the four interacting dark energy models in the $\Omega_m$--$\gamma$ plane.
In these three figures, the 68.3\% and 95.4\% CL posterior distribution contours are shown. The data combinations used are the current only and the current+SL 30-yr combinations, and their constraint results are shown with red and blue contours, respectively. Here we use ``current'' to denote the current SN+BAO+CMB+$H_0$ data combination for convenience.
One can clearly see that the degeneracies are well broken with the SL test data for all the models.

\begin{table*}\tiny
\caption{Errors of parameters in the $w$CDM, I$w$CDM1, I$w$CDM2, I$w$CDM3 and I$w$CDM4 models for the fits to
current only and current+SL 30-yr data.}
\label{table2}
\small
\setlength\tabcolsep{2pt}
\renewcommand{\arraystretch}{1.3}
\begin{tabular}{cccccccccccccccc}
\\
\hline\hline &\multicolumn{5}{c}{current only} &&\multicolumn{5}{c}{current + SL 30-yr} \\
           \cline{2-6}\cline{9-13}
Error  & $w$CDM & I$w$CDM1 & I$w$CDM2 & I$w$CDM3 & I$w$CDM4 && & $w$CDM & I$w$CDM1 & I$w$CDM2 & I$w$CDM3 & I$w$CDM4 \\ \hline
$\sigma(w)$              & $0.082$
                   & $0.083$
                   & $0.096$
                   & $0.092$
                   & $0.099$&&
                   & $0.062$
                   & $0.066$
                   & $0.082$
                   & $0.082$
                   & $0.086$\\
$\sigma(\gamma)$              & $-$
                   & $0.0095$
                   & $0.0300$
                   & $0.0279$
                   & $0.0340$&&
                   & $-$
                   & $0.0066$
                   & $0.0273$
                   & $0.0252$
                   & $0.0310$
                   \\
$\sigma(\Omega_{m})$       & $0.0140$
                   & $0.0142$
                   & $0.0143$
                   & $0.0139$
                   & $0.0139$&&
                   & $0.0026$
                   & $0.0045$
                   & $0.0060$
                   & $0.0058$
                   & $0.0058$\\

$\sigma(H_0)$              & $1.81$
                   & $2.51$
                   & $1.77$
                   & $1.88$
                   & $1.78$&&
                   & $0.64$
                   & $1.81$
                   & $0.99$
                   & $1.04$
                   & $1.01$\\

\hline
\end{tabular}
\end{table*}

The 1$\sigma$ errors of the parameters $w$, $\gamma$, $\Omega_m$, and $H_0$ for the five models with
current only and current+SL 30-yr data are given in Table~\ref{table2}.
With the 30-yr SL observation, the constraints on $\Omega_m$ and $H_0$ will be improved, respectively,
 by 81.4\% and 64.6\% for the $w$CDM model, by 68.3\% and 27.9\% for the I$w$CDM1 model, by 58.0\% and 44.1\% for the I$w$CDM2 model, by 58.3\% and 44.7\% for the I$w$CDM3 model, and by $58.3\%$ and 43.3\% for the I$w$CDM4 model.
Therefore, we can see that with a 30-yr observation of the SL test the geometric constraints on dark energy would be improved enormously.
For all the considered interacting dark energy models, the constraints on $\Omega_m$ and $H_0$ would be improved, relative to the current joint observations, by about 60\% and 30--40\%, with the SL 30-yr observation.

We also discuss the impact of the SL test data on constraining the equation of state $w$ and the coupling $\gamma$.
The SL 30-yr observation helps improve the constraints on $w$ by 24.4\%, 20.5\%, 14.6\%, 10.9\% and 13.1\% for the $w$CDM, I$w$CDM1, I$w$CDM2, I$w$CDM3 and I$w$CDM4 models, respectively. The SL 30-yr observation helps improve the constraints on $\gamma$ by 30.5\%, 9.0\%, 9.7\% and 8.8\% for the I$w$CDM1, I$w$CDM2, I$w$CDM3 and I$w$CDM4 models, respectively.
Therefore, we find that among the four interacting dark energy models, the I$w$CDM1 model is the best one in the sense that the constraint results could be improved by the SL test data.

\begin{figure}
\begin{center}
\includegraphics[width=8cm]{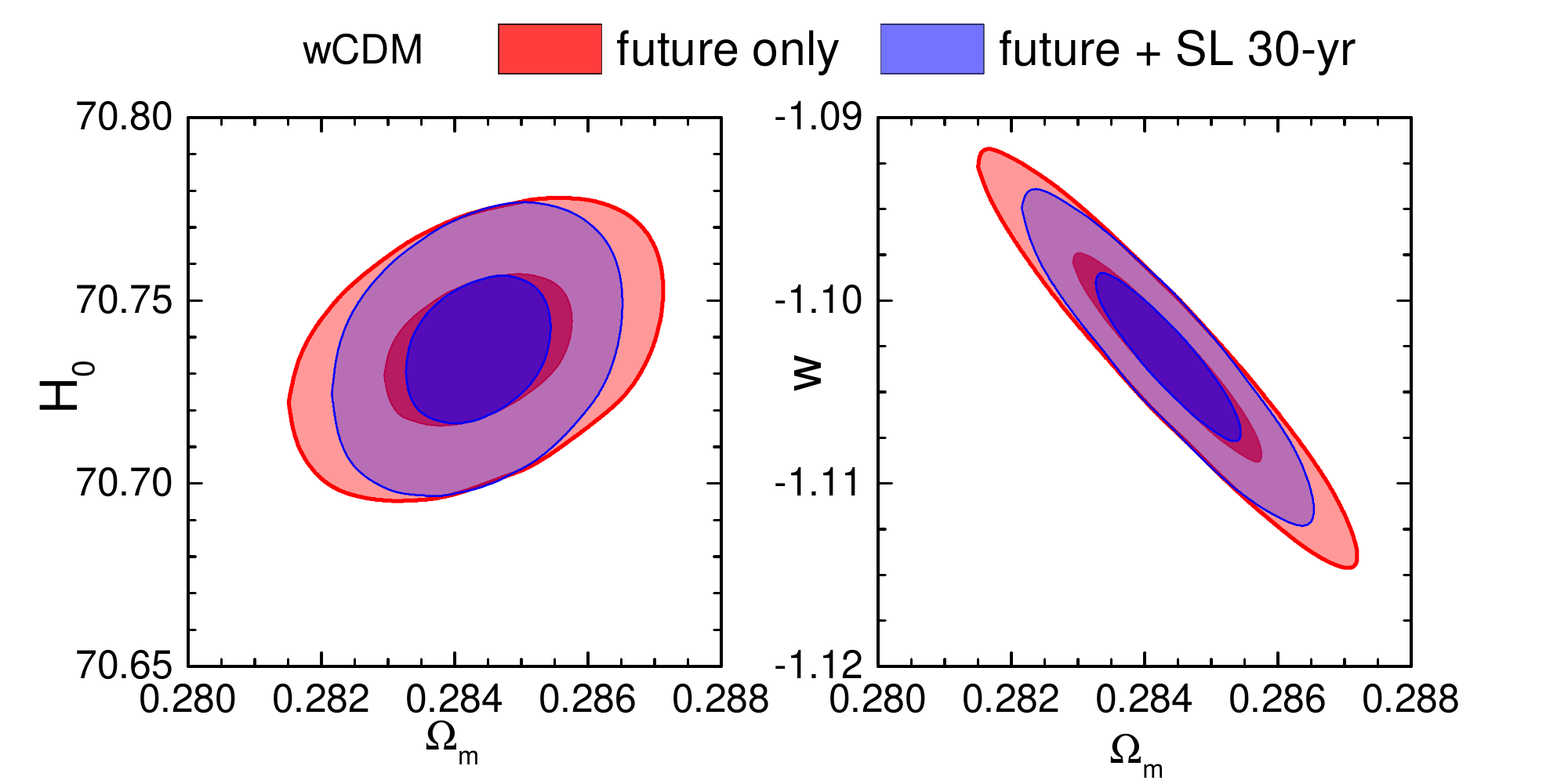}
\end{center}
\caption{Constraints (68.3\% and 95.4\% CL) in the $\Omega_m$--$H_0$ plane and in the $\Omega_m$--$w$ plane for $w$CDM model with future only and future+SL 30-yr data.}
\label{fig4}
\end{figure}

\begin{figure}
\begin{center}
\includegraphics[width=8cm]{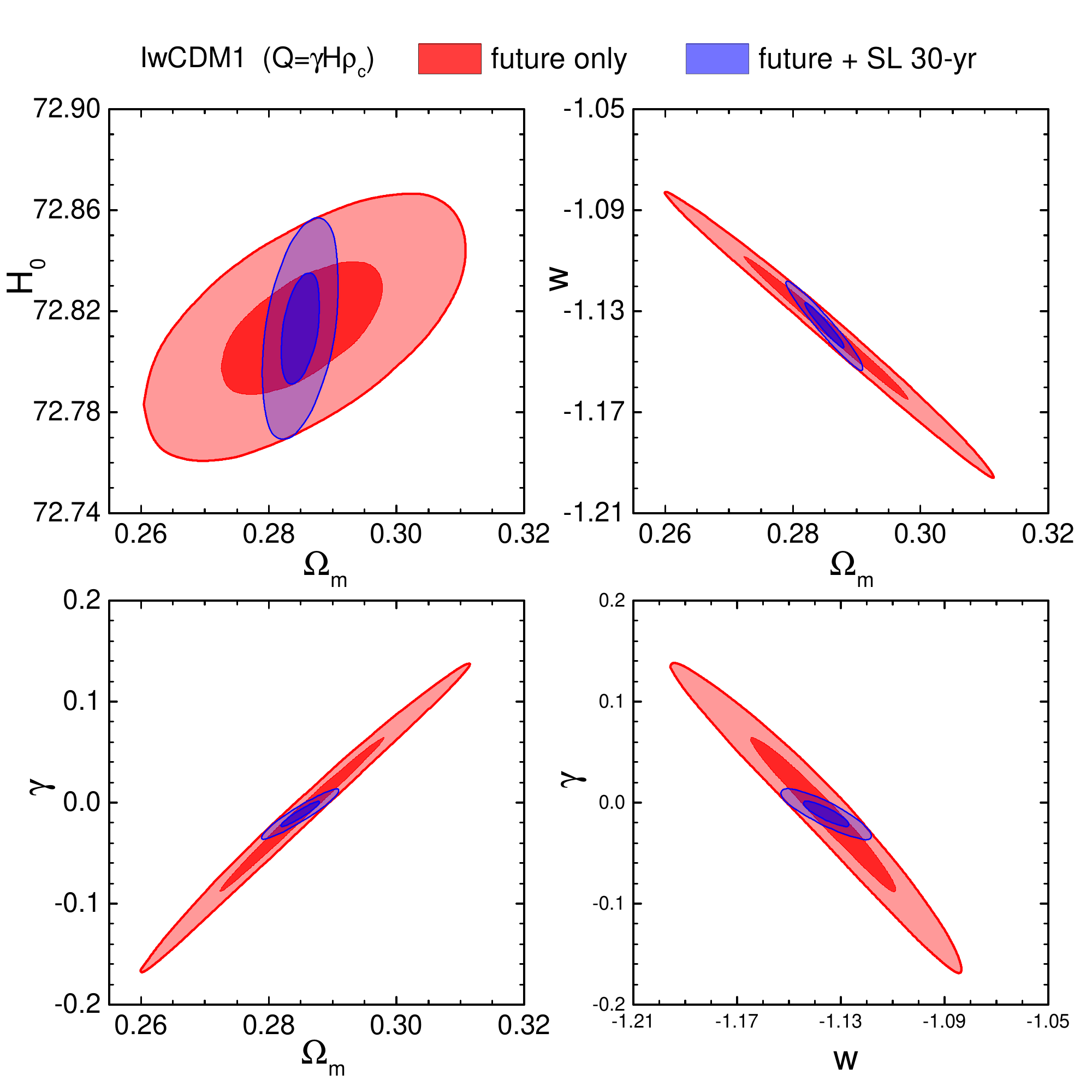}
\end{center}
\caption{Constraints (68.3\% and 95.4\% CL) results for I$w$CDM1 model with future only and future+SL 30-yr data.}
\label{fig5}
\end{figure}

\begin{table}\tiny
\caption{Errors of parameters in the $w$CDM and I$w$CDM1 models for the fits to
future only and future+SL 30-yr data.}
\label{table3}
\small
\setlength\tabcolsep{2pt}
\renewcommand{\arraystretch}{1.3}
\begin{tabular}{cccccccccccccccccccccc}
\\
\hline\hline &\multicolumn{2}{c}{future only} &&\multicolumn{2}{c}{future + SL 30-yr} \\
           \cline{2-3}\cline{5-6}
Error  & $w$CDM & I$w$CDM1  && $w$CDM & I$w$CDM1  \\ \hline
$\sigma(w)$              & $0.0083$
                   & $0.0416$&
                   & $0.0067$
                   & $0.0132$\\
$\sigma(\gamma)$              & $-$
                   & $0.1119$&
                   & $-$
                   & $0.0181$
                   \\
$\sigma(\Omega_{m})$       & $0.0021$
                   & $0.0190$&
                   & $0.0016$
                   & $0.0045$\\

$\sigma(H_0)$              & $0.0296$
                   & $0.0374$&
                   & $0.0286$
                   & $0.0315$\\

\hline
\end{tabular}
\end{table}

We have discussed the quantification of the impact of future SL test data on constraining interacting dark energy from current SN+BAO+CMB+$H_0$ observations. The results show that future SL test data can effectively break the degeneracies in the current data for interacting dark energy models and thus will provide fairly important supplement to the other observations. In the following, we will further explore what role the SL test will play in the future geometric constraints on interacting dark energy. We take the I$w$CDM1 model (with $Q=\gamma H\rho_c$) as an example for this analysis and compare the result with that of the $w$CDM model. As mentioned above, we simulate future SN and BAO data based on the WFIRST mission in this analysis.

Figure~\ref{fig4} shows the joint constraints on the $w$CDM model in the $\Omega_m$--$H_0$ and $\Omega_m$--$w$ planes.
Figure~\ref{fig5} shows the joint constraints on the I$w$CDM1 model in the $\Omega_m$--$H_0$, $\Omega_m$--$w$, $\Omega_m$--$\gamma$, and $w$--$\gamma$ planes, respectively.
The 68.3\% and 95.4\% CL posterior distribution contours are shown. The data combinations used are the future only and the future + SL 30-yr combinations, and their constraint results are shown with red and blue contours, respectively.
The 1$\sigma$ errors of the parameters $w$, $\gamma$, $\Omega_m$, and $H_0$ for the $w$CDM model and the I$w$CDM1 model for the above data combinations are given in Table~\ref{table3}.
Note that here we use ``future'' to denote the data combination of future SN and BAO for convenience.
It is shown that with the 30-yr SL observation, the constraints on $\Omega_m$ and $H_0$ will be improved by
23.8\% and 3.4\% for the $w$CDM model, and by $76.3\%$ and 15.8\% for the I$w$CDM1 model.
For the $w$CDM model, the constraints on $w$ can be improved by 19.3\%, with the SL 30-yr observation, while for the I$w$CDM1 model, the SL 30-yr observation helps improve the constraints on $w$ and $\gamma$ by 68.3\% and 83.8\%, respectively.

Comparing Figs. \ref{fig4} and \ref{fig5}, we find that in the future geometric constraints, the redshift drift observation could not break the degeneracies for the $w$CDM model, but could efficiently break the degeneracies for the I$w$CDM1 model. In the future geometric constraints, for the I$w$CDM1 model, the SL 30-yr observation would help improve the measurement precisions of
$\Omega_m$, $H_0$, $w$ and $\gamma$ by more than 75\%, 15\%, 65\%, and 80\%, respectively.

The main purpose of this work is to investigate the possible impact of future SL test data on existing geometric measurements. In the future, in case $\Lambda$CDM model is excluded, one important work is to constrain the coupling parameter $\gamma$, if we wish to find evidence for the existence of interaction between dark sectors. Therefore, it is quite meaningful to investigate the impact of future SL test data on parameter estimation for interacting dark energy models. For all considered interacting $w$CDM models in this work, it is shown that SL test can effectively break the existing parameter degeneracies and greatly improve the precisions of parameter estimation. The results are consistent with the cases of the $\Lambda$CDM, $w$CDM, and $w_0w_a$CDM  models \cite{msl1, msl2}. By considering more models, we can conclude that the improvement of parameter estimation by SL test data is independent of the cosmological models in the background. This shows the importance of including SL test data in future cosmological constraints.

\section{Summary}

In this paper, we have analyzed how the redshift drift measurement (i.e. SL test signal) would impact on parameter estimation for the interacting dark energy models. By detecting redshift drift in the spectra of Lyman-$\alpha$ forest of distant quasars, SL test directly measures the expansion rate of the universe in the redshift range of $2\lesssim z \lesssim 5$, providing an important supplement to other probes in dark energy constraints. We consider four typical interacting dark energy models: (\romannumeral1) $Q=\gamma H\rho_c$, (\romannumeral2) $Q=\gamma H\rho_{de}$, (\romannumeral3) $Q=\gamma H_0\rho_c$, and
(\romannumeral4) $Q=\gamma H_0\rho_{de}$.

Following our previous works \cite{msl1, msl2}, in order to guarantee that the simulated SL test data are consistent with the other geometric measurement data,
we used the best-fitting dark energy models constrained by the current combined geometric measurement data as the fiducial models to produce the mock SL test data.

We showed that the SL test data are extremely helpful in breaking the existing parameter degeneracies.
Compared to the current SN+BAO+CMB+$H_0$ constraint results, the 30-yr observation of SL test could improve the constraints for all the considered interacting dark energy models on
$\Omega_m$ and $H_0$ by about 60\% and 30--40\%, while the constraints on $w$ and $\gamma$ can be only slightly improved.

We also quantified the impact of SL test data on interacting dark energy constraints in the future geometric measurements. To do this analysis, we simulated the future SN and BAO data based on the long-term space-based project WFIRST. We found that the SL test could also play a crucial role in the future joint geometric constraints, especially for the constraints on $w$ and $\gamma$. Taking the interacting dark energy model with $Q=\gamma H\rho_c$ as an example, the 30-yr observation of
SL test would help improve the measurement precision of $\Omega_m$, $H_0$, $w$ and $\gamma$ by more than 75\%, 15\%, 65\%, and 80\%, respectively.

\begin{acknowledgments}
This work was supported by the National Natural Science Foundation of
China under Grant No.~11175042, the Provincial Department of Education of
Liaoning under Grant No.~L2012087, and the Fundamental Research Funds for the
Central Universities under Grants No.~N140505002, No.~N140506002, and No.~N140504007.
\end{acknowledgments}


\begin{thebibliography}{}

\bibitem{PPF1}
Y.~H.~Li, J.~F.~Zhang and X.~Zhang,
  Phys.\ Rev.\ D {\bf 90}, 063005 (2014)
  [arXiv:1404.5220 [astro-ph.CO]].
%

\bibitem{PPF2}
Y.~H.~Li, J.~F.~Zhang and X.~Zhang,
  Phys.\ Rev.\ D {\bf 90}, 123007 (2014)
  [arXiv:1409.7205 [astro-ph.CO]].

\bibitem{sandage}
A. Sandage,
 Astrophys. J. {\bf 136}, 319 (1962).

\bibitem{loeb}
  A.~Loeb,
  Astrophys.\ J.\  {\bf 499}, L111 (1998)
  [astro-ph/9802122].


  %
%
\bibitem{sl1}
  P.~-S.~Corasaniti, D.~Huterer and A.~Melchiorri,
  Phys.\ Rev.\ D {\bf 75}, 062001 (2007)
  [astro-ph/0701433].

\bibitem{sl2}
  A.~Balbi and C.~Quercellini,
  Mon.\ Not.\ Roy.\ Astron.\ Soc.\  {\bf 382}, 1623 (2007)
  [arXiv:0704.2350 [astro-ph]].

\bibitem{sl3}
  H.~-B.~Zhang, W.~Zhong, Z.~H.~Zhu and S.~He,
  Phys.\ Rev.\ D {\bf 76}, 123508 (2007)
  [arXiv:0705.4409 [astro-ph]].

\bibitem{sl4}
  J.~Zhang, L.~Zhang and X.~Zhang,
  Phys.\ Lett.\ B {\bf 691}, 11 (2010)
  [arXiv:1006.1738 [astro-ph.CO]].


\bibitem{sl5}
  Z.~Li, K.~Liao, P.~Wu, H.~Yu and Z.~-H.~Zhu,
  Phys.\ Rev.\ D {\bf 88}, 2, 023003 (2013)
  [arXiv:1306.5932 [gr-qc]].


\bibitem{sl6}
  S.~Yuan, S.~Liu and T.~-J.~Zhang,
  arXiv:1311.1583 [astro-ph.CO].

\bibitem{sl7}
 M. Martinelli, S. Pandolfi, C. J. A. P. Martins and P. E. Vielzeuf,
 Phys. Rev. D {\bf 86}, 123001 (2012)
 [arXiv:1210.7166 [astro-ph.CO]].

%
%
%
\bibitem{Darling}
J.~Darling,
 Astrophys.\ J.\ Lett.\  {\bf 761}, L26 (2012)
[arXiv:1211.4585 [astro-ph.CO]].

 \bibitem{Zhang21}
H.~-R.~Yu, T.~-.~Zhang and U.~-L.~Pen
Phys. Rev. Lett. {\bf 113}, 041303 (2014)
[arXiv:1311.2363 [astro-ph.CO]].

\bibitem{Liske}
  J.~Liske, A.~Grazian, E.~Vanzella, M.~Dessauges, M.~Viel, L.~Pasquini, M.~Haehnelt and S.~Cristiani {\it et al.},
  Mon.\ Not.\ Roy.\ Astron.\ Soc.\  {\bf 386}, 1192 (2008)
  [arXiv:0802.1532 [astro-ph]].

\bibitem{msl1}
J.~J.~Geng, J.~F.~Zhang and X.~Zhang,
  JCAP {\bf 07}, 006 (2014)
  [arXiv:1404.5407 [astro-ph.CO]].

\bibitem{msl2}
 J.~J.~Geng, J.~F.~Zhang and X.~Zhang,
  JCAP {\bf 12}, 018 (2014)
  [arXiv:1407.7123 [astro-ph.CO]].

 \bibitem{cosmomc}
A. Lewis and S. Bridle,
Phys. Rev. D {\bf 66}, 103511 (2002)
 [astro-ph/0205436].

\bibitem{snls3}
  A.~Conley {\it et al.}  [SNLS Collaboration],
  Astrophys.\ J.\ Suppl.\  {\bf 192}, 1 (2011)
  [arXiv:1104.1443 [astro-ph.CO]].

\bibitem{6dF}
  F.~Beutler, C.~Blake, M.~Colless, D.~H.~Jones, L.~Staveley-Smith, L.~Campbell, Q.~Parker and W.~Saunders {\it et al.},
    Mon.\ Not.\ Roy.\ Astron.\ Soc.\  {\bf 416}, 3017 (2011)
    [arXiv:1106.3366 [astro-ph.CO]].


\bibitem{DR7}
  N.~Padmanabhan, X.~Xu, D.~J.~Eisenstein, R.~Scalzo, A.~J.~Cuesta, K.~T.~Mehta and E.~Kazin,
   Mon.\ Not.\ Roy.\ Astron.\ Soc.\  {\bf 427}, no. 3, 2132 (2012)  [arXiv:1202.0090 [astro-ph.CO]].

\bibitem{DR9}
  L.~Anderson, E.~Aubourg, S.~Bailey, D.~Bizyaev, M.~Blanton, A.~S.~Bolton, J.~Brinkmann and J.~R.~Brownstein {\it et al.},
   Mon.\ Not.\ Roy.\ Astron.\ Soc.\  {\bf 427}, no. 4, 3435 (2013)  [arXiv:1203.6594 [astro-ph.CO]].  

\bibitem{WiggleZ}
  C.~Blake, S.~Brough, M.~Colless, C.~Contreras, W.~Couch, S.~Croom, D.~Croton and T.~Davis {\it et al.},
  Mon.\ Not.\ Roy.\ Astron.\ Soc.\  {\bf 425}, 405 (2012)  [arXiv:1204.3674 [astro-ph.CO]].  




\bibitem{WW}
  Y.~Wang and S.~Wang,
  Phys.\ Rev.\ D {\bf 88}, 043522 (2013)
   [arXiv:1304.4514 [astro-ph.CO]].


%
\bibitem{Riess2011}
A. G. Riess {\it et al.},
 Astrophys.\ J.\  {\bf 730}, 119 (2011)
 [arXiv:1103.2976 [astro-ph.CO]].


\bibitem{DETF}
  A.~Albrecht, G.~Bernstein, R.~Cahn, W.~L.~Freedman, J.~Hewitt, W.~Hu, J.~Huth and M.~Kamionkowski {\it et al.},
  astro-ph/0609591.  
 %
%


%
%

%
\end{thebibliography}
\end{document}